\documentstyle[11pt,epsf]{article}
\def\r{\rightarrow}
\def\l{\lambda}
\def\p{\psi}
\def\bc{\begin{center}}
\def\ec{\end{center}}

\begin{document}

\rightline{\sf{hep-ph/0003227}}

\bc
{\Large\bf {New Physics Effects from B Meson Decays} }\footnote{Talk
presented at the Sixth Workshop on High Energy Particle Phenomenology
(WHEPP-6), Chennai, India}
\\[1cm]
{\large\bf Anirban Kundu}\\[0.3cm]
E-mail: akundu@juphys.ernet.in\\
Department of Physics, Jadavpur University, Calcutta - 700032, India
\ec

\vspace{1cm}
\abstract{
In this talk, we point out some of the present and future 
possible signatures of physics beyond the Standard Model from 
B-meson decays, taking R-parity conserving and 
violating supersymmetry as illustrative examples.}

\vspace{2cm}

\section{Introduction}

It has long been established that the B-meson system (both charged
and neutral) may be the ideal place to look for indirect effects
of physics, both CP-conserving and CP-violating,
beyond the Standard Model (BSM) \cite{reviews}. 
The reasons, in brief, are:
\begin{itemize}
\item The $B-\bar B$ mixing is dominated by the short-distance box
diagram with the top quark running inside the loop. Thus, CP-violation
is large and the major part of it, fortunately, can be calculated
to a good precision. The long-distance part is anyway known to be
negligible in B-decays (this may be compared with D decays where it is
the dominant contribution).                          
\item
The soft QCD effects are less pronounced for B than for D --- this
is due to the $m_c/m_b$ suppression. Thus, $1/m_b$ corrections are
at least controllable.
\item Due to the CKM suppression, the lifetime of the B meson is 
sufficiently large to be accurately measured. This has important implications
in the asymmetric B-factories.
\item Due to the abovementioned reasons, we will have a number of
dedicated B-factories in the near future, apart from the running ones like
CLEO. We will need the hadronic machines to measure CP-violation
in modes with branching ratios (BR) less than $10^{-5}$. 
\end{itemize} 

However, before one proceeds, one must remember that the theoretical 
uncertainties are still significant, and will probably remain so in near 
future \cite{quinn}. 
These uncertainties are dominated by our ignorance of soft-QCD
physics. For example, approximations like factorization and quark-hadron
duality (both local and global) are not at all beyond doubt; 
the numerical
inputs like the strange quark mass, the number of effective colour $N_c$ 
and the regularization scale $\mu$ are not yet certain and should be treated 
as more or less free parameters. With all these handicaps, 
it is {\em extremely} difficult to find the signature of 
BSM physics if that is more than one order of magnitude smaller than the 
SM contribution. 

Fortunately, there are cases when the BSM signal may be equally (or more)
large as the SM one, and can be easily distinguished.   
There are two major ways to proceed.
First, one can look for CP-asymmetries, both direct and mixing-induced, and
see whether they tally with the SM predictions. 
Thus, if the SM amplitude for a particular process be $A_{SM}\exp(i\theta_
{SM})$ and the BSM amplitude for the same process be $A_{BSM}\exp(i\theta_
{BSM})$, the total amplitude is given by
\begin{equation}
A_{tot}e^{i\theta_{tot}}=A_{SM}e^{i\theta_{SM}}(1+he^{i\phi})
\end{equation}
where $h\equiv A_{BSM}/A_{SM}$ and $\phi=\theta_{BSM}-\theta_{SM}$. The change
in CP-asymmetry is essentially governed by $h$, which should be $\sim
{\cal O}(1)$ for the BSM physics to be visible. 
Such investigations involve the
measurement of the angles as well as the sides of the unitarity triangle
(UT). Here, one may face a number of different situations, some of 
which are:\\
(i) The three angles of the UT do not sum up to $\pi$. This is a definite
signal of new physics, but, considering the errors in determining the 
angles, may not be easily obtained even in the B-factories. Even if one
measures the functions $\sin 2\alpha$, $\sin 2\beta$ and $\sin^2\gamma$
from CP-asymmetries, one needs to resolve the discrete ambiguity to get
the actual values of the angles \cite{kayser}.  \\
(ii) The angles do sum up to $\pi$, but the sides are not in the proper
ratio. One needs to determine the sides too for this type of signal.\\
(iii) CP-asymmetries measured from different modes, which should yield 
the same angle in SM, give different results. For example, $J/\psi K_S$
and $\phi K_S$ modes may produce different CP asymmetries (both should 
give the same angle $\beta$ in SM), and one may find nonzero CP-asymmetries
in $b\rightarrow c$ decay modes of $B_s$ (which, in SM, should not give
any significant CP-asymmetry).\\  
(iv) One can observe sizable asymmetries in leptonic, semileptonic and
radiative B-decays too.

Secondly, one can concentrate on CP-conserving observables. A good place 
is the branching ratios (BR) of rare modes. CLEO already has some 
interesting signals
\cite{cleo} which are listed in Table 1; there may be 
more in near future. Another excellent channel is to look for forbidden
modes in the SM (like $B^+\r K^+ K^+ \pi^-$ \cite{huitu}) where even 
a single event may signal BSM physics. OPAL has looked for such signals
and placed limits on BSM couplings \cite{opal}.  

\begin{table}[htbp]
\caption{Branching ratios ($\times 10^6$) of the $\eta K$ and $\eta' K$
modes. The experimental results are at $2\sigma$. }
\bc
\begin{tabular}{|c|c|c|}
\hline
Mode & Theoretical BR & Experiment \\
\hline
$B^+\r \eta' K^+$ & 7-41 & $80^{+10}_{-9}\pm 8$ \\
$B^0\r \eta' K^0$ & 9-33 & $88^{+18}_{-16}\pm 9$ \\
$B^+\r \eta K^{*+}$ & 0.03-9 & $27.3^{+9.6}_{-8.2}\pm 5.0$ \\
$B^0\r \eta K^{*0}$ & 0.05-3 & $13.8^{+5.5}_{-4.4}\pm 1.7$ \\
\hline
\end{tabular}
\ec
\end{table}

Anyway, we should realize that quantification of BSM physics is something 
we must approach with caution; qualitative signals 
({\em e.g.}, $h\sim 1$) are what we can hope to
observe quickly. Of course, if BSM physics is indicated from other
experiments, then the B-system can be used to complement and quantify
that.

\section{Basic Formalism}

\begin{figure}[htbp]
\epsfxsize=8cm
\centerline{\epsfbox{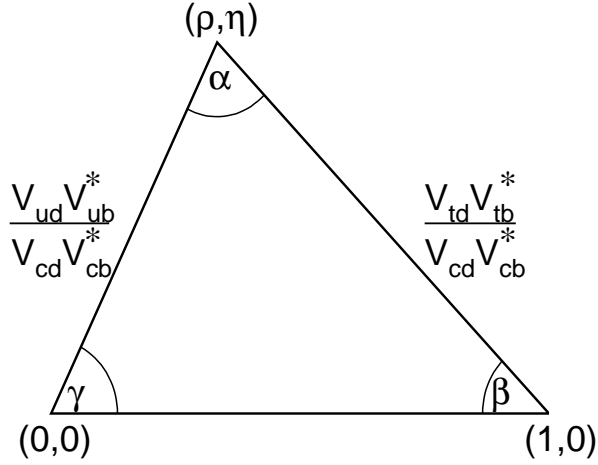}}
\caption{The Unitarity Triangle.}
\label{fig:ut}
\end{figure}

The unitarity triangle probed in B-decays is given by the orthogonality
condition
\begin{equation}
V_{ub}V_{ud}^* + V_{cb}V_{cd}^* + V_{tb}V_{td}^* = 0.
\end{equation}
The triangle, alongwith the angles $\alpha$, $\beta$ and $\gamma$, is shown
in Fig. \ref{fig:ut}. 
The only two complex entries in the CKM matrix in Wolfenstein 
parametrization (WP) are
\begin{equation}
V_{td} = |V_{td}| exp(-i\beta),\ \ \ V_{ub} = |V_{ub}| exp(-i\gamma).
\end{equation}
With WP, the tip of the UT has coordinates $(\rho,\eta)$.
The central values are \cite{parodi2000}
\begin{equation}
\rho(1-\lambda^2/2) = 0.240^{+0.057}_{-0.047},\ \ 
\eta(1-\lambda^2/2) = 0.335 \pm 0.042.
\end{equation}


In the SM, the quark-level subprocesses that are important to
determine the angles of the UT are shown in Table 2, which is
mainly taken from \cite{buras}. 
It is helpful to remember that $B^0-
\bar{B^0}$ mixing measures $2\beta$, $b\rightarrow u$ measures $2\gamma$,
presence of both simultameously measures $2\alpha$ (assuming the UT 
closes), and
$B_s-\bar{B_s}$ mixing and $b\rightarrow c$ decay are CP-conserving to
a very good extent.
Some of such CP-conserving modes are also
shown; a nonzero CP-asymmetry in them (say, in $B_s\r J/\psi\phi$)
would be an encouraging signal for BSM physics.

\begin{table}[htbp]
\caption{Quark-level subprocesses for B-decays. P and V denote pseudoscalar
and vector mesons respectively.}

\bc
\begin{tabular}{|l|c|c|c|l|}
\hline
No& Quark level& Type& Meson level&  Remarks\\
\hline
  &            &     &            &       \\
1&$b\r d\bar u u$ & $P_1P_2$ & $\bar{B^0}\r\pi^+\pi^-$ & $\alpha$
(penguin pollution)\\
\hline
2&$b\r d\bar c c$ & $P_1P_2$ & $\bar{B^0}\r  D^+  D^-$ & $\beta$
(clean)\\
3&$b\r d\bar c c$ & $P  V  $ & $\bar{B^0}\r J/\p\pi^0$ & $\beta$
(penguin pollution)\\
4&$b\r d\bar c c$ & $P  V  $ & $\bar{B_s}\r J/\p  K_S$ & $\lambda^2
\eta$ (very clean)\\
\hline
5&$b\r s\bar u u$ & $P_1P_2$ & $\bar{B^0}\r\pi^0 K_S $ & $\alpha,
\gamma$ (not so clean) \\
6&$b\r s\bar u u$ & $P_1P_2$ & $\bar{B_s}\r  K^+ K^- $ & $\gamma$
(clean)\\
\hline
7&$b\r s\bar c c$ & $P_1P_2$ & $\bar{B_s}\r D_s^+D_s^-$ & $\lambda^2\eta$
(very clean)\\
8&$b\r s\bar c c$ & $P  V  $ & $\bar{B^0}\r J/\p  K_S$ & $\beta$
(gold-plated)\\
9&$b\r s\bar c c$ & $V_1V_2$ & $\bar{B_s}\r J/\p \phi $ & CP-conserving
\\
\hline
10&$b\r d\bar s s$ & $P_1P_2$ & $\bar{B^0}\r K^0\bar{K^0}$ &
QCD penguin dominates\\
11&$b\r d\bar s s$ & $P  V  $ & $\bar{B^0}\r \pi^0 \phi $ &
EW penguin dominates\\
\hline
12&$b\r s\bar s s$ & $P  V  $ & $\bar{B^0}\r K_S \phi $ & $\beta$
(clean)\\
\hline
13&$b\r s\bar d d$ & $P_1P_2$ & $\bar{B_s}\r K^0\bar{K^0}$ &
QCD penguin\\
\hline
14a&$b\r u\bar c s$ & $P_1P_2$ & $  B^- \r \bar{D^0} K^-$ & $DK$
triangles \\
14b&$b\r c\bar u s$ & $P_1P_2$ & $  B^- \r     {D^0} K^-$ & measure
$\gamma$ \\
\hline
\end{tabular}
\ec
\end{table}

\section{Possible New Physics}

In this section, we first briefly review a couple of non-SUSY extensions of
the SM, and the results are taken mainly from \cite{gronau}. After that
we discuss two versions of SUSY. 

\subsection{Four Generations}

With four quark generations, the CKM matrix is $4\times 4$, with three 
independent phases. This makes UT a quadrangle, and the asymmetries 
measured by different processes will be different from their SM 
predictions: for example, the asymmetry measured in $B^0-\bar{B^0}$
mixing is not only $2\beta$ but some $2(\beta+\theta_d)$ due to the $t'$
mediated box.

The smoking gun signals may be the simultaneous measurements of
$\alpha$, $\beta$ and $\gamma$ which will not sum up to $\pi$. Also,
$b\rightarrow d\gamma$ and $b\rightarrow d\ell^+\ell^-$ may be 
enhanced compared to their SM values depending on the magnitudes of
$V_{td}$ and $V_{t'd}$ \cite{hou}. CP asymmetry in $B\r J/\psi K_S$ 
is negative for almost half of the parameter space, and almost 40\%
of the parameter space predicts the magnitude of CP asymmetry in
$B_s\r J/\psi~\phi$ to be more than $0.2$ (the SM asymmetry is almost
zero) \cite{london}.

\subsection{Multi-Higgs Doublet with no FCNC}

In such models, the CP-asymmetries are almost identical to that of the
SM, since the CKM matrix still has the same structure, and
$H^+\bar{u_i} d_j$ couplings have the same phase as that of the SM.
There may be a significant change in
the total amplitude of $B^0-\bar{B^0}$ mixing due to the
$H^+$ box diagrams, which will in turn
affect the value of $V_{td}$.

An interesting signal in this model may be the $B^0\rightarrow \ell^+\ell^-$
rates, which, for some particular choice of the parameter space, can be
much higher than the SM ones.

We do not discuss the spontaneous CP-violation scenario, since only
spontaneous CP-violation would mean a real CKM matrix, which is ruled out
from the $K_L-K_S$ mass difference and the CDF measurement of $\sin
2\beta$.

\subsection{Supersymmetry with R-parity Conservation}

The minimal SUSY and its R-parity conserving variants are interesting
mainly for the CP-violating observables; any CP-conserving observable
like the BRs must have two SUSY particles in the loop and is thereby
suppressed in general.

As is well-known, there can be two independent phases in the minimal SUSY
which can lead to interesting CP-violating effect. They can be written as
\begin{equation}
\phi_A = arg(A^*m_{1/2}), \ \ \ \phi_B = arg (m_{1/2}\mu (m_{12}^2)^*),
\end{equation}
where the symbols have their usual meanings. For both these phases $\sim 
{\cal O}(1)$, the dipole moment of neutron, for example, is larger than
the experimental limit by two to three orders of magnitude for 100 GeV
squarks. This is known as the Supersymmetric CP problem. Another problem
arises in the measurement of $\epsilon_K$ which turns out to be seven orders
of magnitude larger than the experimental number unless there is some
fine-tuning among the parameters. 

To solve the SUSY flavour problems regarding $\epsilon_K$ and dipole
moment of neutron, a number of different models were proposed. Among them
are: (1) Heavy squarks at the TeV scale; (2) Universality among right
and left squark masses for different generations; (3) Alignment of quark
and squark mixing matrices, and (4) Approximate CP-symmetry of the
Lagrangian.
There are a number of specific
flavour models in the literature which incorporates
one or more of the above features \cite{barenboim}. As has been pointed out,
the effects on measured observables crucially depend on the exact structure
of the model, and not all models in a given category have same
CP-violating predictions.
For example, alignment-type models predict $\sim {\cal O}(1)$
contributions to $B^0-\bar{B^0}$ mixing phases but very small contribution
to the neutron dipole moment; heavy squark models (for the first two
generations) may have a larger $d_n$.
In Table 3, which is taken from \cite{nir}, we summarise the predictions
of various type of models. For a detailed discussion, see 
\cite{barenboim,nir}.

\begin{table}[htbp]
\caption{Prediction of different SUSY flavour models. $\theta_d$ is the
phase change from the SM prediction $\beta$ in $B^0-\bar{B^0}$ mixing.
$a$ denotes CP-asymmetries for the two decay channels. Taken from
{\protect\cite{nir}}. }
\bc
\begin{tabular}{|c|c|c|c|c|}
\hline
Model & $d_n/d_n^{exp}$ & $\theta_d$ & $a_{D^0\r K^-\pi^+}$
& $a_{K\r \pi\nu\bar\nu}$ \\
\hline
SM & $\leq 10^{-6}$ & 0 & 0 & ${\cal O}(1)$ \\
Exact & $\leq 10^{-6}$ & 0 & 0 & $\approx$ SM \\
Universality & & & & \\
Approx. & $\geq 10^{-2}$ & ${\cal O}(0.2)$ & 0 & $\approx$ SM \\
Universality & & & & \\
Approx. CP & $\sim 10^{-1}$ & $-\beta$ & ${\cal O}(10^{-3})$
& ${\cal O}(10^{-5})$ \\
Alignment & $\geq 10^{-3}$ & ${\cal O}(0.2)$ & ${\cal O}(1)$ & $\approx$ SM \\
Heavy $\tilde q$ & $\approx 10^{-1}$ & ${\cal O}(1)$ & ${\cal O}(10^{-2})$ &
$\approx$ SM \\
\hline
\end{tabular}
\ec
\end{table}

Another interesting observable is the forward-backward lepton asymmetry
(as well as the absolute BRs) in $B\r X_s\ell^+\ell^-$ where $\ell=e$ or $\mu$
\cite{scimemi}. For both the leptons, the SM predictions for $A_{FB}$ is
$0.23$ but it can vary from $0.33$ to $-0.18$ in SUSY models. The negative
$A_{FB}$ constitutes an interesting signal. The BRs can be enhanced by a factor
of four or can be suppressed by a factor of two, which should also be
measured in the B-factories.

\subsection{Supersymmetry without R-parity}

R-parity violating (RPV) SUSY has one great advantage over the non-RPV
SUSY models: the new physics contributions appear in the tree-level, 
and hence can greatly enhance or suppress the SM contributions. Here we
will discuss the popular approach, {\em i.e.}, we will consider all RPV
couplings to be free parameters, constrained only by various experimental
data, and study its consequences on B-decays. 

The physics of RPV has been discussed in detail in the literature, and 
here we will only quote the main results. 
For $B\r M_1M_2$ decays ($M$ is any meson in general) the relevant pair
of couplings 
is either $\lambda'\lambda'$ or $\lambda''\lambda''$ type (but not both
simultaneously). For $B\r M\ell^+{\ell'}^-$ decays, it is $\lambda\lambda'$
and $\lambda'\lambda'$ together. For example, we can have sneutrino/squark 
mediated $b\r d_i\bar{d_j} d_k$ decays and selectron/squark mediated $b\r d_i
\bar{u_j} u_k$ decays ($i,j,k$ are generation indices). All 
B-decay modes are affected by suitable pair of
RPV couplings; more specifically, all UT angles
can change from their SM predictions. In SM, the decays $B\r J/\psi K_S$ 
and $B\r \phi K_S$ measure the same angle $\beta$; with RPV, the measured
CP-asymmetries may be different, which will definitely signal new physics
\cite{guetta}. One can see forbidden modes like $B^+\r K^+K^+\pi^-$ 
originating from the SM forbidden $b\r s s \bar d$ decay \cite{huitu}. 
CP-asymmetries $\sim 100\%$ in the measurement of the UT angle $\gamma$
can be obtained even from $B^+$ decays \cite{datta};
thus, study of $B^+$s, alongwith $B^0$s and $B_s$s,
are of paramount importance.
The leptonic forward-backward asymmetries are modified too: for a pure
$\lambda\lambda'$ type coupling, there is no FB asymmetry, whereas for a 
$\lambda'\lambda'$ type coupling, it is in the opposite direction from
SM \cite{jang}.

\begin{figure}[htbp]
\epsfxsize=8cm
\centerline{\epsfbox{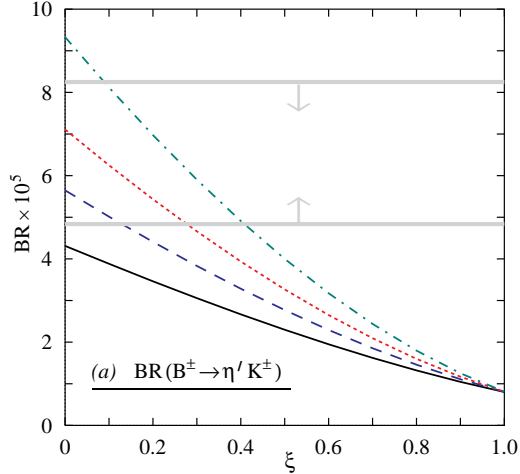}}
\caption{BR for $B\r \eta'K$ as a function of $\xi$. The solid curve
gives the SM value. In the presence of a $\lambda'_{i23}\lambda'_{i22}$
operator with $m_{\tilde\nu}=200$ GeV, the long-dashed, short-dashed
and dot-dashed curves correspond to the cases where each of the two
$\lambda'$s equal 0.05, 0.07 and 0.09 respectively. The thick lines
correspond to the experimental bounds.}
\label{fig:etaprk}
\end{figure}

Another important feature is that RPV couplings can enhance or suppress
the BRs significantly. As we have seen, the CLEO data on $\eta' K$ and
$\eta K^*$ are quite far away from the SM prediction. It has been 
shown \cite{choudhury} that a moderate value of the product coupling
$d^R_{222}\equiv\lambda'_{i23}\lambda'_{i22}$
(each $\lambda'=0.05$-$0.09$, say), perfectly compatible with the 
experimental bounds, can enhance the BRs to their experimental value. 
This is demonstrated in Fig. \ref{fig:etaprk} 
where we have plotted the BR against
$\xi=1/N_c^{eff}$, using $m_s(1~GeV)=165$ MeV. At the same time, this 
product suppresses decays like $B^0\r\phi K$; in SM, this decay is allowed 
only for $\xi<0.23$ but with RPV, larger ranges of $\xi$ are allowed. 

However, the SM range is in conflict with other PV modes such as
$B^{\pm}\r \omega K^{\pm}$ and $B^{\pm}\r \omega\pi^{\pm}$. The former
requires either $\xi<0.05$ or $0.65<\xi<0.85$ while the latter requires
$0.45<\xi<0.85$ \cite{deshpande}.  
Interestingly, the $d^R_{222}$ operator affects $B^\pm\r \phi K^\pm$
while the other two decay modes are blind to it. This
additional contribution interferes destructively with the SM amplitude and
$BR(B^\pm\r \phi K^\pm)$ is suppressed leading to a wider allowed range
for $\xi$. For example, with each $\lambda'=0.09$,
$\xi$ can be as large as $0.8$, thus allowing for a common
fit to all the three ($PV$) modes under
discussion\footnote{Note that the favoured value of $\xi$ for the $PP$
        and $PV$ modes still continue to be different. While this is {\em not}
        a discrepancy, a common $\xi$ for both these sets can  be
        accommodated for values of $\l'$ slightly larger than that we
        have considered.}.
(Note that $\lambda'=0.09$ is just ruled out from $B\r X_s\nu\bar\nu$ data
\cite{grossman,aleph} if we assume a sneutrino-squark degeneracy; but there
is every reason to believe that squarks are heavier than sneutrinos, in which
case our analysis still stands.)
$d^R_{222}$ also affects $VV$ decay modes such as
($B\rightarrow\phi K^*$). As this calculation involves a few more
model  dependent parameters, we do not discuss it here.

\section{Conclusions}

The study of B-decays, both in the CP-conserving and CP-violating fronts,
is quite interesting to study indirect effects of new physics, more so in
view of the upcoming B-factories. CLEO has already given some food for
thought. Among various new physics models, non-SUSY extensions of the SM
mainly affect the $B^0-\bar{B^0}$ amplitude, and, maybe, phase. The 
determination of $V_{td}$ may be affected too. Different SUSY flavour models 
will have different signatures regarding the neutron dipole moment and
asymmetries in $J/\psi K_S$ and $\pi\nu\bar\nu$ modes. RPV SUSY models
can contribute to almost all B-decays, and can even induce some SM forbidden
decyas. One of the important achievements is to explain the CLEO result
on $B\r\eta'K$ with RPV. 

However, this is only about the observation of BSM physics, 
and to have a qualitative
measurement, one needs to minimize the theoretical errors, which will be
the biggest challenge to the theoreticians in the next few years. In short,
we await some really exciting years on both theoretical and experimental
fronts! 

\section{Acknowledgements}

I thank Rahul Basu and other organisers of WHEPP-6 for providing a most
stimulating atmosphere. Some of the material presented here is based on
the work done with Debajyoti Choudhury and Bhaskar Dutta.


\end{document}